\documentclass[a4paper,11pt]{article}
\usepackage{pos}
\usepackage{graphicx}
\usepackage{xcolor}
\usepackage{amsmath}
\usepackage{physics}
\usepackage{ragged2e}
\usepackage{caption}
\usepackage{subcaption}
\usepackage[numbers]{natbib}
\newcommand{\bt}{\mathbf{b}}
\newcommand{\qso}{Q_{\mathrm{s},0}} 
 
\newcommand{\rt}{\mathbf{r}}

\newcommand{\kt}{\mathbf{k}}

\newcommand{\xbj}{x_\mathrm{Bj}}
\newcommand{\nf}{n_\mathrm{f}}
\newcommand{\seq}{\sum_{q}^{\nf} e_q^2}

\newcommand{\btheta}{\boldsymbol{\theta}}
\newcommand{\ymodel}{\mathbf{y}(\btheta)}

\newcommand{\yexp}{\mathbf{y}_{\mathrm{exp}}}
\newcommand{\as}{\alpha_\mathrm{s}}
\newcommand{\der}{\mathrm{d}}

\renewcommand{\logo}{\relax}

\title{Bayesian approach to determine the initial condition for the Balitsky-Kovchegov equation}

\author*[a,b]{Carlisle Casuga}
\author[a,b]{Mikko Karhunen}
\author[a,b]{Heikki Mäntysaari}

\affiliation[a]{University of Jyväskylä,  P.O. Box 35, 40014 University of Jyväskylä, Finland}

\affiliation[b]{Helsinki Institute of Physics, P.O. Box 64, 00014 University of Helsinki, Finland}

\emailAdd{carlisle.doc.casuga@jyu.fi}
\emailAdd{mikko.a.karhunen@student.jyu.fi}
\emailAdd{heikki.mantysaari@jyu.fi}

\abstract{We present posterior distributions of parameters that characterize the nonperturbative initial input for the Balitsky-Kovchegov evolution equation. The BK equation evolves an initial dipole-target scattering amplitude at moderate $\xbj$ values toward smaller values. We use Bayesian inference to constrain the model parameters against the precise combined reduced cross section data from HERA, and obtain probability distributions for the free parameters. Our results show the data's preference for the anomalous dimension to be $\gamma \sim 1$. The distributions provide a rigorous method for propagating uncertainties of the BK initial condition for other CGC calculations. We demonstrate this for inclusive quark production in proton-proton collisions and nuclear modification factor for the total deep inelastic scattering cross section to be measured at the EIC.}

\FullConference{31st International Workshop on Deep Inelastic Scattering (DIS2024)\\
 8–12 April 2024\\
Grenoble, France\\}

\begin{document}
\maketitle

\section{Introduction}

Results from the future Electron-Ion Collider (EIC) \cite{AbdulKhalek2021} must be met with high precision theoretical predictions in order to probe saturation and proton structure at small-$x$. The color glass condensate (CGC) effective field theory \cite{Gelis2010} provides a convenient framework to describe scattering in the high energy regime. Along with the next-to-leading corrections for the CGC cross sections, global analysis of the nonperturbative input to evolution equations like the Balitsky-Kovchegov (BK) equation ~\cite{Kovchegov1999,Balitsky1996} is necessary to move towards higher theoretical precision. The BK initial condition has been fitted to HERA data before in Refs. \cite{Albacete2009, Lappi2013, Albacete2011, Ducloue2020} at the leading order and Refs. \cite{Beuf2020, Hanninen2023} at the next-to-leading order. This study \cite{Casuga2024} also aims to constrain the BK initial condition. Contrary to previous fits, our analysis provides uncertainties for the parameter estimates for the first time. This input is necessary for statistically rigorous propagation of uncertainties when calculating observables in CGC.

In the dipole picture of deep inelastic scattering (DIS), the virtual photon fluctuates into a quark-antiquark pair and interacts with the proton target modeled as a CGC field. The lifetime of the dipole is much larger than the time scale of its interaction with the proton. One can then factorize the transverse (T) and longitudinal (L) photon-proton cross section as
\begin{equation}
\sigma^{\gamma^* p}_{T,L}= \sigma_0 \sum_f \int \der^2 \rt \int \frac{\dd z}{4\pi}|\psi^{\gamma^* \to q\bar q}(\rt,Q^2,z)|^2 N(\rt,\xbj).
\end{equation}
Here $\psi^{\gamma^* \to q\bar q}(\rt,Q^2,z)$ describes the $\gamma^* \rightarrow q\bar{q}$ splitting at photon virtuality $Q^2$ and the longitudinal momentum fraction carried by the quark, $z$. The amplitude, $N(\rt,\xbj)$, describes the dipole-target scattering at a particular dipole size, $\rt$, and Bjorken-$x$, $\xbj$. The BK equation evolves a nonperturbative initial amplitude towards small $\xbj$ values or higher rapidities, $Y = 1/ \ln \xbj$. For the initial condition, we use a McLerran-Venugopalan model \cite{Mclerran1994} inspired parametrization as in Ref. \cite{Lappi2013},
\begin{equation}
        N(\rt, x=x_0) = 1-\exp \left[  - \frac{\left(\rt^2 \qso^2\right)^\gamma}{4} \ln \left( \frac{1}{|\rt| \Lambda_\text{QCD}} +  e_c\cdot  e   \right)   \right]. 
\end{equation}

The parameters describing the initial condition are the following: the parameter that controls $Q_s$, $\qso^2$, the anomalous dimension, $\gamma$, and infrared regulator, $e_c$. We introduce the transverse proton size, $\sigma_0/2$, another free parameter to replace impact parameter dependence such that $\int \mathrm{d}^2\bt \rightarrow \sigma_0/2$ . The running of the strong coupling in the transverse coordinate space is parametrized as $\as \sim (\ln (C^2))^{-1}$ where the $C^2$ is a parameter that controls the speed of evolution of the BK kernel. Our study focuses on finding the probability distribution of these parameters by constraining against the most recent combined inclusive deep inelastic $ep$ scattering data from HERA \cite{H12015} in the region 2.0 $\mathrm{GeV}^{2} < Q^2 < $ 50.0 $ \mathrm{GeV}^2$. Finally, our analysis neglects heavy quarks and is limited to the leading order (impact factor and BK equation). 

\section{Bayesian Inference}

We find posterior distributions for the parameters using Bayesian inference setup. In this work, we use a similar procedure as in other works in the field used i.e. in extracting properties of the quark-gluon plasma \cite{Bernhard2016, Parkkila2022} and event-by-event fluctuating proton geometry \cite{Mantysaari2022}. The workflow involves a combination of a gaussian process emulator (GPE) and a Monte Carlo Markov Chain (MCMC) sampler. Because the MCMC sampler requires a computationally efficient model surrogate, this study uses a GPE \cite{Pedregosa2011} in place of the model. The emulator only needs training from model calculations across points in the parameter space to learn the parameter dependence of the model. It represents each predictive point as a gaussian process, characterized by a mean and standard deviation value, by fitting a covariance kernel with hyperparameters over the training points. 

The MCMC importance sampler is an algorithm that employs a random walk across the parameter space finding areas where the model and data values are closest. These are areas of high posterior. Bayes' theorem defines the posterior function as
\begin{equation}
      P(\ymodel | \yexp) \propto P(\yexp | \ymodel) P(\btheta),
\end{equation}
where $ P(\yexp | \ymodel)$ is the likelihood function and $P(\btheta)$ is the prior at a certain parameter point $\btheta$. The likelihood is a function of the difference between the model calculation at certain $\btheta$ and the experimental value, taking into account experimental and model uncertainties. Meanwhile, the prior, as used in this study, is a flat distribution that spans the parameter space that encodes \textit{prior} knowledge about possible values or range of the parameters. In the study, the MCMC sampling is done with \texttt{emcee} \cite{Foreman2013} that have multiple random walkers, whose positions are dependent on each other, exploring the parameter space. The acceptance probability of a step of a walker is the ratio of the posterior function between the proposed and the current step. This way the walkers will eventually converge to areas of high posterior and their chain will form samples of the posterior distribution. 

\section{Results}

\begin{figure}[ht]
    \centering
    \includegraphics[width=\linewidth]{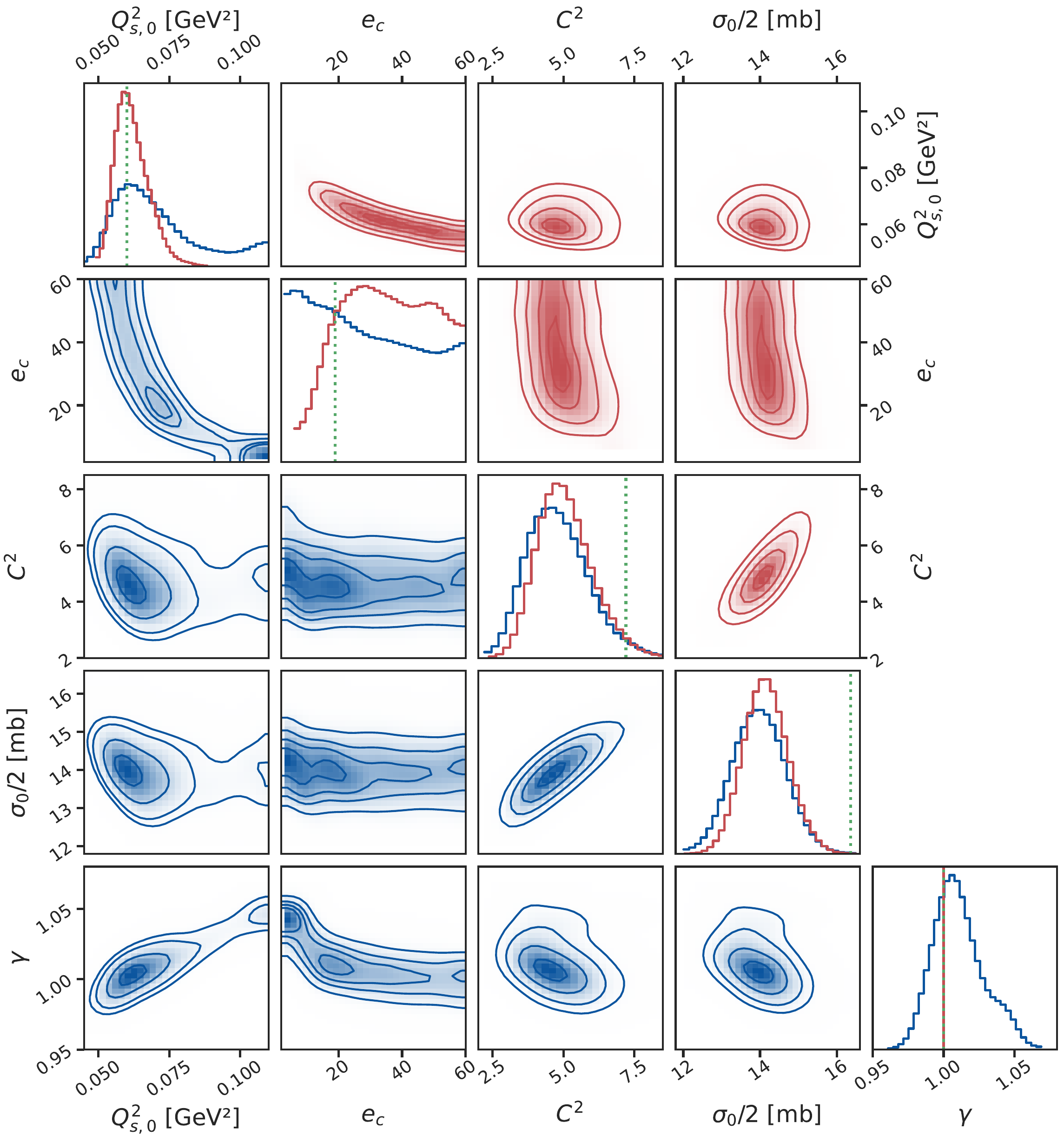}
    \caption{Posterior distributions for both the 4-parameter (red) and 5-parameter (blue) case. The dotted green line shows best fit values from $\mathrm{MV}^e$ parametrization of Ref. \cite{Lappi2013}}
    \label{posteriordistribution}
\end{figure}

The five-dimensional posterior distribution for $\btheta = (\qso^{2}, \gamma, e_{c}, C^{2}, \sigma_0/2)$ obtained through Bayesian analysis are the blue plots shown in Figure \ref{posteriordistribution}. Moreover, the red plots present another setup where we fix $\gamma = 1$.  The diagonal histograms represent the probability distributions for each parameter.  The distributions for most likely values of the parameters have values close to those obtained in previous leading order dipole fits \cite{Albacete2009, Lappi2013, Albacete2011, Ducloue2020}. We also get a very good agreement with the HERA data, obtaining $\chi^2/\mathrm{dof} = 1.02$ averaged over many samples for both the 4- and 5-parameter setups. Correlated systematic uncertainties from the experimental data are included in this analysis, the effect of which are wider posterior distributions but unchanged correlations between parameters. 

We find well-constrained parameters except $e_c$. We expected that the dipole amplitude is not very sensitive to $e_c$ especially for small dipoles. In the setup where $\gamma$ is a free parameter, the  posterior distribution is wider and the anomalous dimension covers values of $\gamma = 0.95 ... 1.05$. This is a favorable result as values $\gamma \geq 1$ produce negative 2DFT values hence negative cross sections \cite{Giraud2016} for inclusive quark production. 

The off-diagonal plots inform the correlations between each pair of parameters, for example, the negative correlation between $\qso^2$ and $\sigma_0/2$. In the region where the contribution from small dipoles dominate, the dipole-target amplitude is $N(r) \sim (\qso^2/Q^2)^\gamma$ where $\sigma^{\gamma^*p} \sim \frac{\sigma_0}{2} N(r)$. This also explains the correlation between $\gamma$ and $\sigma_0/2$. 

\begin{figure*}[t!]
    \centering
    \begin{subfigure}[t]{0.49\textwidth}
        \centering
        \includegraphics[width = \textwidth]{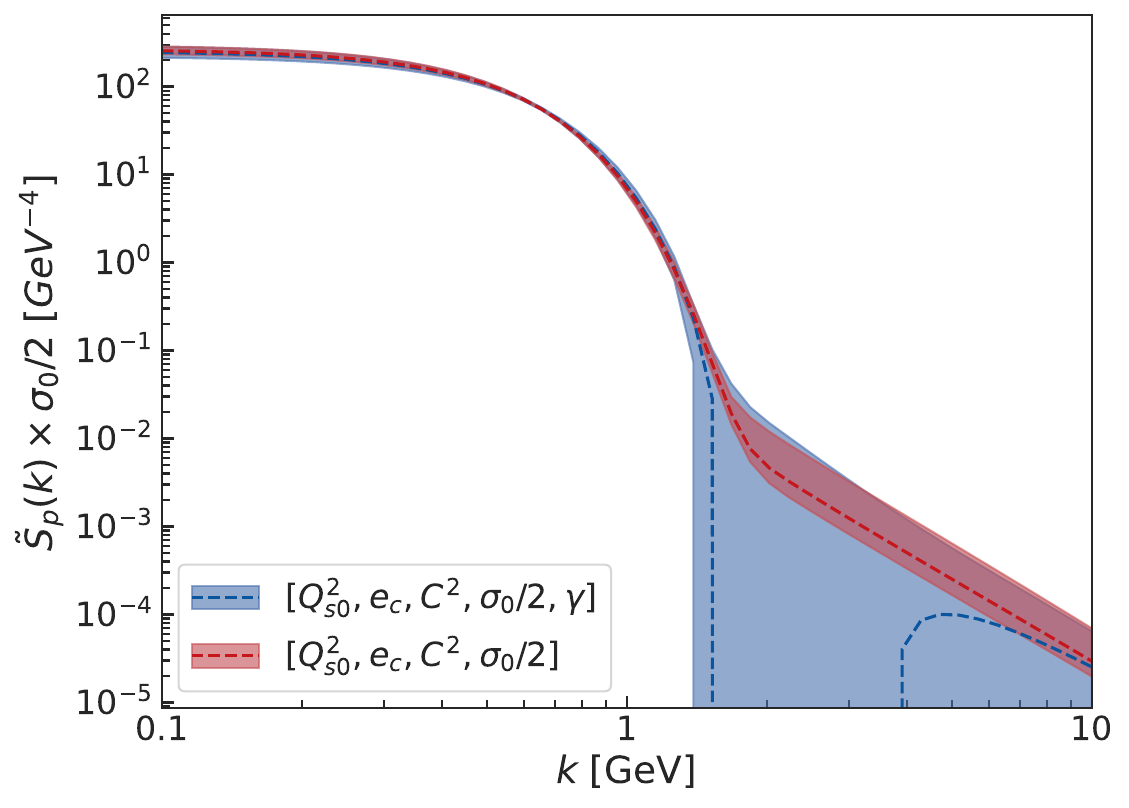}
        \caption{}
        \label{sps}
    \end{subfigure}%
    ~ 
    \begin{subfigure}[t]{0.49\textwidth}
        \centering
        \includegraphics[width = \textwidth]{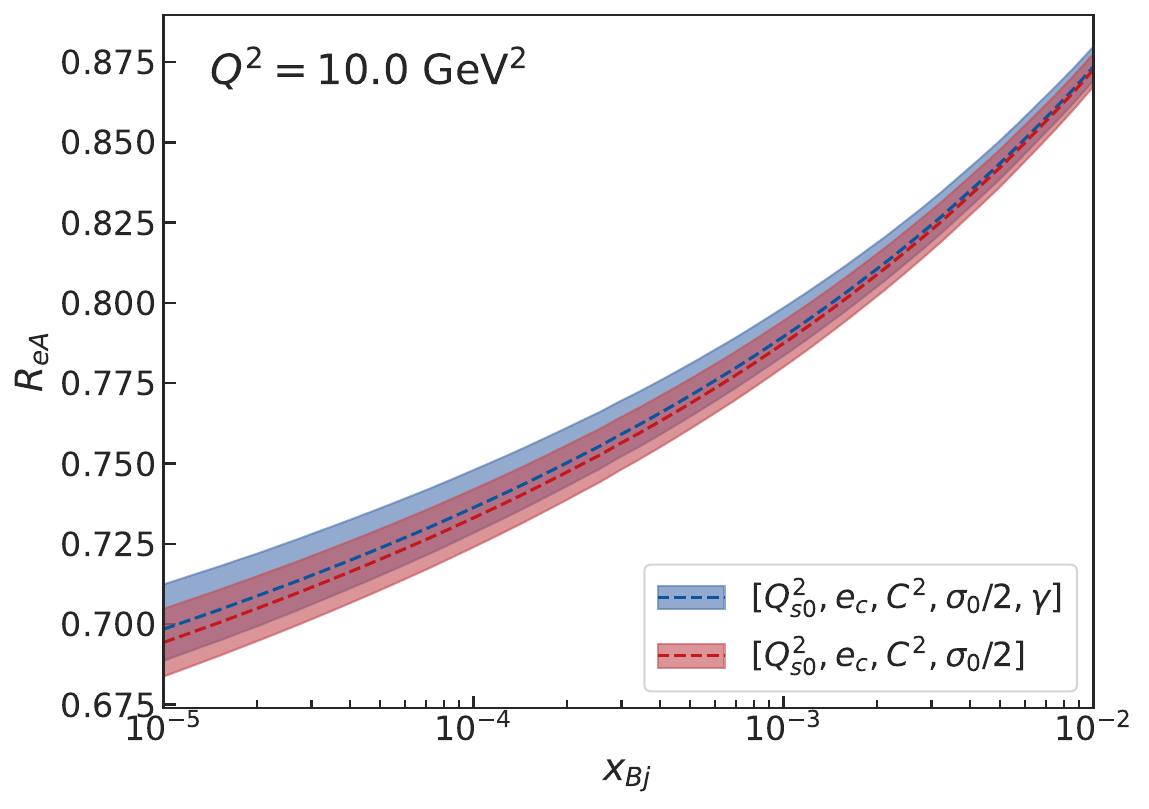}
        \caption{}
        \label{rea}
    \end{subfigure}
    \caption{Applications of posterior distributions: (a) 2DFT of the dipole-proton amplitude scaled by $\sigma_0/2$ at initial $\xbj$ in the 4-parameter (red) and 5-parameter (blue) cases. (b) Nuclear modification factor, $R_{eA}$ for the $F_2$ of as a function of $\xbj$ at fixed $Q^2 = 10.0$ $\mathrm{GeV}^2$ and $A = 197$.}
\end{figure*}

The posterior distributions provide a streamlined method of propagating uncertainties to observables that can be calculated in the CGC framework. We demonstrate explicit examples by calculating single inclusive forward quark production in proton-proton collisions from samples of the posterior. It is proportional to the 2-dimensional Fourier transform of the initial dipole-proton amplitude
\begin{equation}
    \tilde S_p(\kt) = \int \dd[2]{\rt} e^{-i\kt \cdot \rt} \left[ 1-N(\rt)\right], 
\end{equation}
shown in Figure \ref{sps}. The results illuminate the importance of the $\gamma \leq 1$ constraint as there are negative values for the 2DFT of the dipole amplitude due to $\gamma \geq 1$ values in the posterior for the 5-parameter setup.

Figure \ref{rea} shows the nuclear modification ratio, $R_{eA} = F_{2,A}/(AF_{2,p})$, for the structure function, $F_2$. We calculate it for an electron-Au collision for which the $R_{eA}$ will be measured at the EIC. The initial condition for the dipole-nucleus scattering amplitude at a fixed impact parameter is related to the transverse thickness function, $T_A(\bt)$, which is the unintegrated Woods-Saxon potential modeling the distribution of nucleons in the nucleus as in Ref. \cite{Lappi2013}. Unlike the case for the inclusive quark production cross section where the $2\sigma$ uncertainties could reach $\sim 150\%$ the mean value, the uncertainties effectively cancel for the modification factor. 

\section{Summary \& Outlook}

We have presented results from the Bayesian inference of the nonperturbative initial condition for the BK evolution equation. Our fits agree very well with the constraining HERA data. We also successfully account for correlated systematic uncertainties from the experimental data. Contrary to previous fits of the initial condition, our results provide uncertainties to parameter estimates enabling a rigorous method to propagate uncertainties for CGC calculations with examples we calculated in the paper.

The Bayesian setup allows easy extension to next-to-leading order analysis and towards constraining the initial condition for the NLO BK equation. For a global analysis approach, we will also include heavy quark data as an additional constraint to the parameters in the future. 

\acknowledgments

We acknowledge funding support from the Vilho, Yrj\"o, and Kalle V\"ais\"al\"a Foundation, the Research Council of Finland, and the Center of Excellence in Quark Matter and projects 338263 and 346567. This work is also supported under the European Union’s Horizon 2020 research and innovation programme by the European Research Council (ERC, grant agreements No. ERC-2018-ADG-835105 YoctoLHC, and ERC-2023-101123801 GlueSatLight) and by the STRONG-2020 project (grant agreement No. 824093). This work used computing resources from the Finnish Grid and Cloud Infrastructure (persistent identifier \texttt{urn:nbn:fi:research-infras-2016072533}). The content of this article does not reflect the official opinion of the European Union and responsibility for the information and views expressed therein lies entirely with the authors.

\bibliographystyle{JHEP}
\bibliography{refs.bib}

\end{document}